\documentclass[reprint, amsmath, amssymb, aps, prl]{revtex4-2}

\usepackage{graphicx}% Include figure files
\usepackage{dcolumn}% Align table columns on decimal point
\usepackage{bm}% bold math
\usepackage{hyperref}% add hypertext capabilities
\usepackage{physics}
\usepackage{xfrac}
\usepackage{xcolor}
\usepackage{soul}

\newcommand{\comment}[1]{}

\begin{document}

\preprint{APS/123-QED}

\title{Hybrid Trapping of Cold Atoms with Surface Forces and Blue-Detuned Evanescent Light on a Nanophotonic Waveguide}% Force line breaks with \\

\author{Riccardo Pennetta}
\email{riccardo.pennetta@hu-berlin.de}
\author{Antoine Glicenstein}
\author{Philipp Schneeweiss}
\author{J\"urgen Volz}
\author{Arno Rauschenbeutel}
\email{arno.rauschenbeutel@hu-berlin.de}

\affiliation{%
 Department of Physics, Humboldt Universit\"at zu Berlin, 12489 Berlin, Germany\\
}%

\date{\today}% It is always \today, today, but any date may be explicitly specified

\begin{abstract}

We demonstrate a novel hybrid nanophotonic trap for cold neutral atoms, leveraging surface forces for attraction and blue-detuned evanescent light for repulsion. We attribute the attractive potential to a combination of Casimir-Polder interactions and electrostatic charges distributed on the waveguide surface. Despite the trap's shallow depth, we efficiently load atoms into it via adiabatic transfer from a conventional two-color dipole trap. Remarkably, the hybrid trap supports a long atomic storage time of 140(9) ms and exhibits a Ramsey coherence time of 16.8(2) ms, the latter exceeding significantly previous reports for nanophotonic systems. Our results pave the way for further exploration of atom-surface interactions at the nanoscale and illustrate the potential of harnessing surface forces to enhance storage and coherence times for atoms coupled to nanophotonic waveguides. This advancement offers new opportunities for neutral-atom quantum technologies.

\end{abstract}

\maketitle

\begin{figure*}[]
\includegraphics[width=0.95\linewidth]{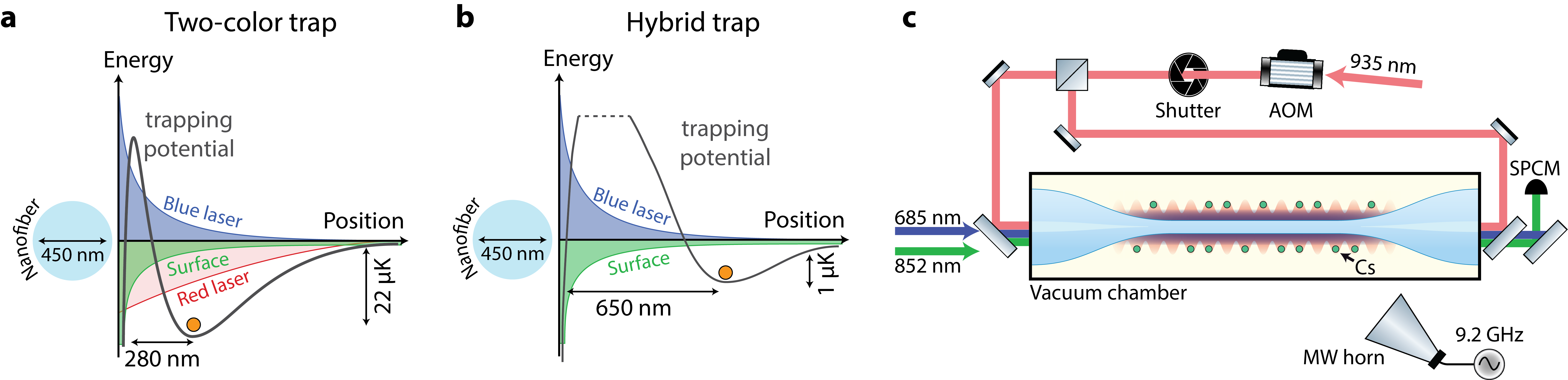}
\caption{\label{fig:FigureSketch} (a, b) Sketch of the potential energy of atoms in (a) a standard nanofiber-based trap and (b) a hybrid trap. (c) Experimental setup for our nanofiber-based cold atom interface. AOM, acousto-optic modulator; Cs, cesium; SPCM, single photon counting module; MW, microwave. 
}
\end{figure*}

Interfacing trapped neutral atoms with light propagating in nanophotonic waveguides enables efficient and long-range photon-mediated atom–atom interactions, opening new frontiers in both cavity quantum electrodynamics (QED) \cite{Reiserer2015} and waveguide QED \cite{Chang2018, Sheremet2023}. These systems are particularly promising in quantum information processing \cite{Dordevic2021} and the generation of non-classical states of light \cite{Prasad2020}, offering a scalable platform for future quantum networks \cite{Menon2024} and exploring many-body quantum phenomena \cite{Liedl2024, Zhou2024a}. A successful approach in this context involves trapping atoms within the evanescent field surrounding an optical nanofiber using a two-color dipole trap \cite{Vetsch2010, Goban2012}. This platform has enabled a variety of applications in, e.g., quantum memories \cite{Corzo2019, Sayrin2015}, chiral light–matter interactions \cite{Lodahl2017} and the study of collective effects \cite{Pennetta2022, Sheremet2023, Liedl2024}. Typically, in this configuration, a red-detuned laser is employed to attract atoms towards the waveguide, while a blue-detuned laser repels them, in order to prevent collisions with the surface. In this way, a stable trapping position can be obtained a few hundred nanometers from the nanofiber surface (see Fig.~\ref{fig:FigureSketch}(a)).

Beyond optical forces, atoms near nanophotonic waveguides are also subject to surface-induced forces, as for instance Casimir–Polder (CP) interaction, which arise from quantum vacuum and thermal fluctuations \cite{Buhmann2012, Schneeweiss2012}, and electrostatic effects due to surface charges \cite{Ong2020, Ocola2024}.
Recent theoretical work has proposed to exploit these surface forces to replace one or both trapping beams \cite{Hung2013, Chang2014a, GonzalezTudela2015a, Chang2018, Kien2022}. Experimentally realizing such schemes would not only deepen our understanding of surface–atom interactions but also mitigate or completely suppress issues typically associated with conventional nanophotonic traps for cold atoms. For example, differential light shifts due to the trapping fields can cause inhomogeneous broadening and degrade the coherence properties of trapped atoms \cite{Reitz2013}. So far, demonstrating surface-force-based traps has remained elusive due to significant challenges, such as the precise nanofabrication required for certain proposed designs \cite{GonzalezTudela2015a} and the typically shallow potential depths of such traps, which complicate direct loading from laser-cooled atomic clouds.

In this work, we experimentally demonstrate a hybrid nanophotonic trap for cold cesium (Cs) atoms, in which the attractive force originates entirely from atom–surface interactions, while a blue-detuned laser provides the repulsive force. In our experiment, we attribute the surface interaction to a combination of the Casimir–Polder interaction and an attractive force resulting from charges on the nanofiber surface. We estimate that the hybrid trap has a depth of $\sim$1 $\mu$K and its trap minimum is located $\sim$650 nm from the nanofiber surface. Remarkably, we observe that Cs atoms in the hybrid trap exhibit longer trap storage times than those in our standard two-color dipole trap, despite the significantly shallower potential. Furthermore, the atoms in the hybrid trap display Ramsey coherence times that exceed by over an order of magnitude those of traditional nanofiber-based traps \cite{Reitz2013} and other nanophotonic systems \cite{Dordevic2021}.
\begin{figure}[]
\includegraphics[width=0.95\linewidth]{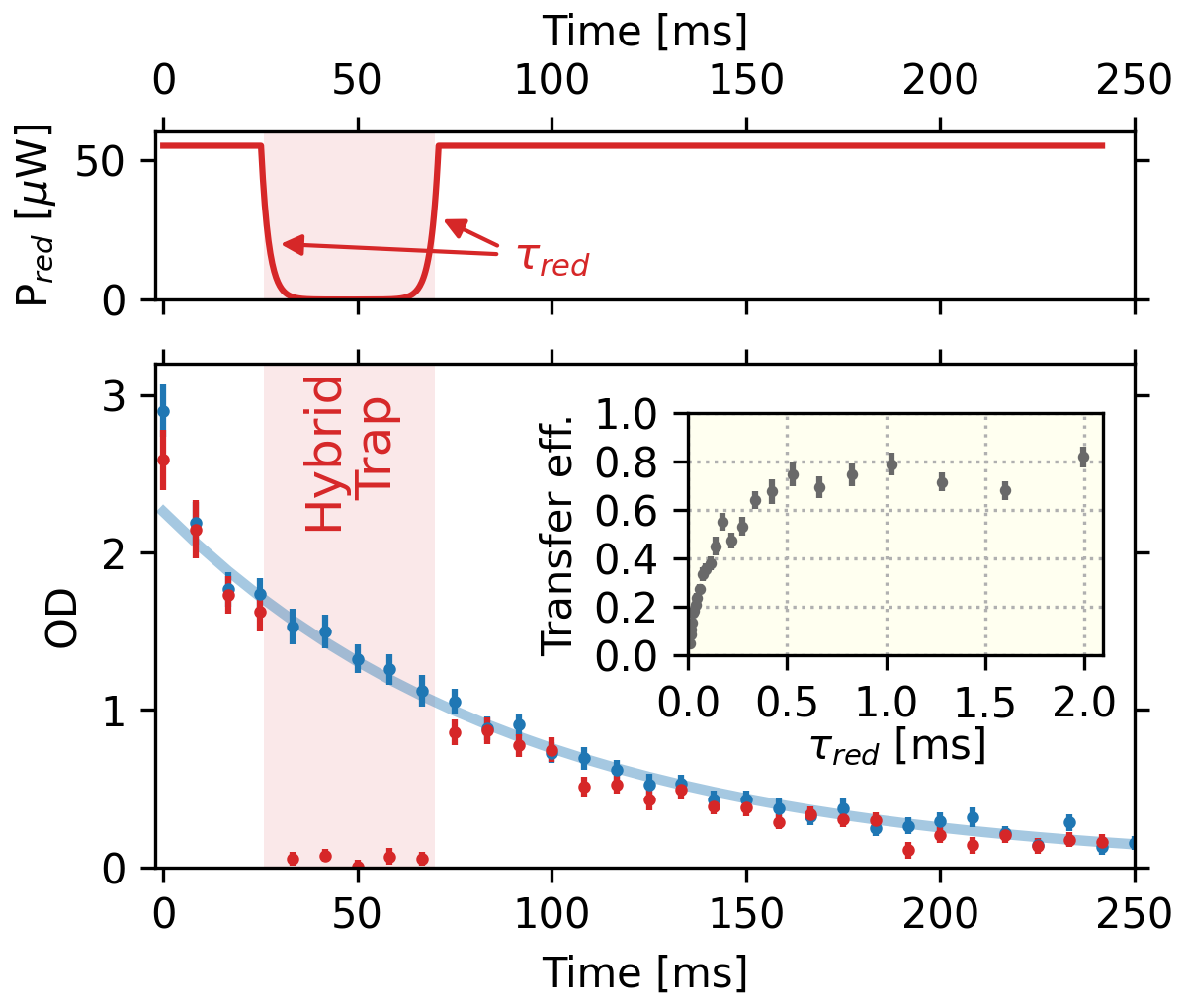}
\caption{\label{fig:FigureLoading}
Bottom panel: the blue dots show a measurement of the optical depth (OD) as a function of holding time for the atoms in our standard two-color dipole trap. The solid blue line is an exponential fit of the data. The red dots refer to a situation in which the power of the red-detuned laser, $P_\text{red}$, is switched off in the interval between 25 ms and 70 ms (i.e. the red-shaded area), thereby transferring the atoms from our standard trap to the hybrid trap. In this case, the atoms are much farther away from the nanofibre surface and therefore measured OD is significantly reduced. After 70 ms, $P_\text{red}$ is ramped up again and the atoms are transferred back into the standard trap. Top panel: $P_\text{red}$ as a function of delay time in this second situation. 
Inset: measured transfer efficiency from the standard trap to the hybrid trap and back as a function of the timescale, $\tau_\text{red}$, with which $P_\text{red}$ is ramped down.
}
\end{figure}

A sketch of the experimental setup is shown in Fig.~\ref{fig:FigureSketch}(c). We optically trap and interface an ensemble of laser-cooled Cs atoms with the evanescent field surrounding an optical nanofiber. The latter is realized as the waist of a tapered optical fiber (TOF) and has a nominal diameter of $2r_0$ = 450 nm, where $r_0$ is the nanofiber radius. A two-color dipole trap is realized using nanofiber-guided light by launching laser beams through the TOF that are blue- and red-detuned with respect to both the Cs D1 and D2 lines. In our experiment, these trapping light fields operate at magic  wavelengths of 935 nm and 685 nm, which do not affect the transition frequency of the Cs D2 transition \cite{Goban2012}. The red-detuned laser is launched from both ends of the TOF to form a standing wave at the nanofiber waist. Its power, $P_\text{red}$, can be controlled using an acousto-optic modulator (AOM) and completely switched off using a mechanical shutter (fall-time $<$500~$\mu$s). By employing optical powers of $P_\text{blue}$ = 8.1 mW and $P_\text{red}$ = 55 $\mu$W, we realize a trap with potential minimum located about 280 nm from the nanofiber surface and a depth of about 22 $\mu$K. In the following, this configuration is referred to as standard trap (cf. Fig.~\ref{fig:FigureSketch}(a)).
Our experiment begins by overlapping the nanofiber with a cloud of Cs atoms in a magneto-optical trap. This is followed by a molasses phase, during which the magnetic fields are switched off and the temperature of the atoms is further reduced via polarization gradient cooling. The temperature of the atoms in the molasses is measured to be 5.5(2) $\mu$K via time-of-flight. During this phase, the trapping sites around the nanofiber are loaded probabilistically, with each site containing at most one atom due to collisional blockade \cite{Schlosser2002}.

The hybrid nanophotonic trap is realized by retaining only the blue-detuned field, which provides a repulsive potential, while the red-detuned beam is completely switched off using the AOM and the shutter. 
We found that directly loading the hybrid trap during the molasses phase is very inefficient due to the shallow depth of the trap.  
Instead, we first load atoms into the standard trap and subsequently transfer them into the hybrid trap by gradually reducing the power of the red-detuned laser, $P_\text{red}$, until it is fully turned off. If this process is performed adiabatically, that is, slowly enough that the change in trap frequency is small over the time duration of a single oscillation period, the atomic motional state populations remain unchanged while the potential landscape is modified \cite{Kastberg1995, Alt2003}.

To demonstrate the transfer process, we present in the bottom panel of Fig.~\ref{fig:FigureLoading} a measurement of the optical depth (OD), measured when probing the atomic ensemble through the TOF with light that is resonant with the D2 transition, as a function of the holding time of the atoms in the trap. The blue dots correspond to measurements where the atoms are held in the standard trap and the blue solid line represents a fit using an exponential function. The red dots show the same measurement, but between 25 ms and 70 ms, the atoms are transferred into the hybrid trap by ramping $P_\text{red}$ down and up adiabatically. Since the atoms in the hybrid trap are significantly farther away from the nanofiber surface, the OD measured through the TOF drops sharply during this interval. To confirm that the atoms remain trapped, we subsequently switch the red-detuned beam back on and observe that the OD nearly returns to its original value. This indicates that most atoms have been successfully transferred to the hybrid trap and back to the standard trap.

To quantify the efficiency of this transfer process, we measure the atom losses after completing the transfer first into the hybrid trap and then back into the standard trap. In this case, we ramp $P_\text{red}$ down and up exponentially with time constant $\tau_{red}$. The inset of Fig.~\ref{fig:FigureLoading} shows the measured transfer efficiency as a function of $\tau_{red}$. If the atom transfer is too fast, the change in potential energy is non-adiabatic, leading to a low transfer efficiency. For longer $\tau_{red}$, the efficiency increases and saturates at 74(4)$\%$ for $\tau_{red} > 500$~$\mu$s. We attribute the residual losses to atoms that initially reside in higher-energy states in the standard trap. For those states that do not have counterparts in the shallower hybrid trap, the atoms are then lost during the transfer. We also tested alternative ramp profiles that might better satisfy the adiabaticity criterion \cite{Alt2003}. However, we did not obtain any significant improvement in efficiency or characteristic timescale.

\begin{figure}[]
\includegraphics[width=0.95\linewidth]{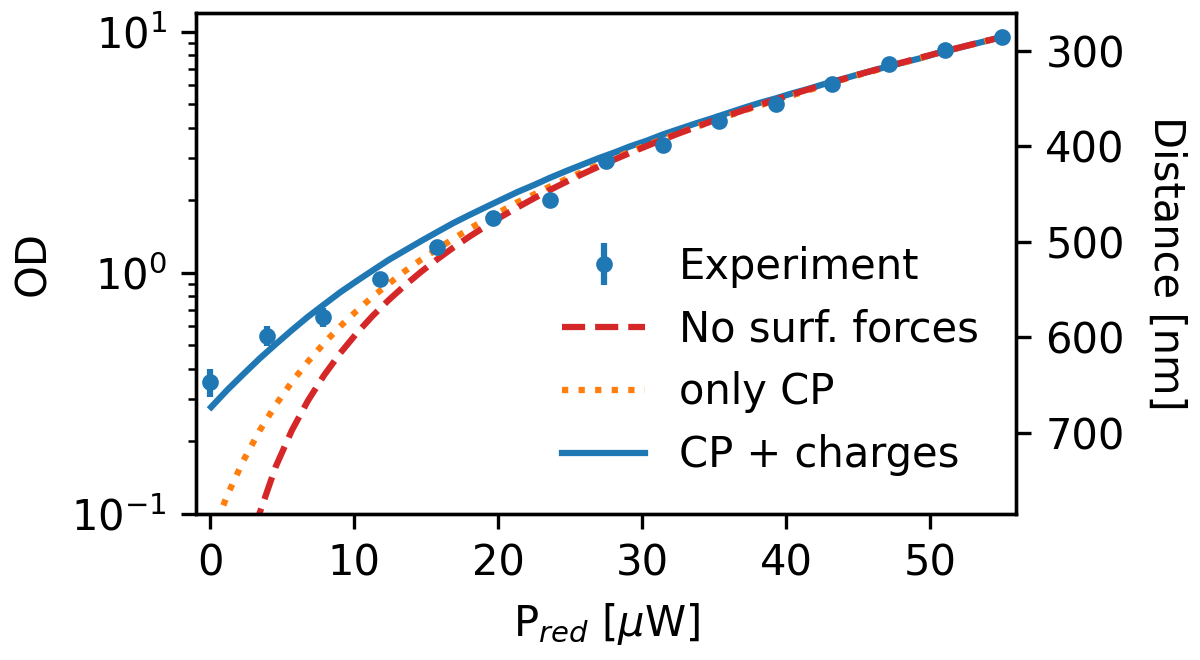}
\caption{\label{fig:FigurePotential}
Measured optical depth (OD) as a function of red-detuned laser power $P_{red}$ (blue dots). The red dashed line shows the calculated OD assuming the atoms experience only the optical trapping potential. The orange dotted line includes an additional surface potential from Casimir–Polder (CP) interaction, while the blue solid line includes both CP interaction and an attractive potential from surface charges (CP + charges).
}
\end{figure}

\begin{figure}[]
\includegraphics[width=0.95\linewidth]{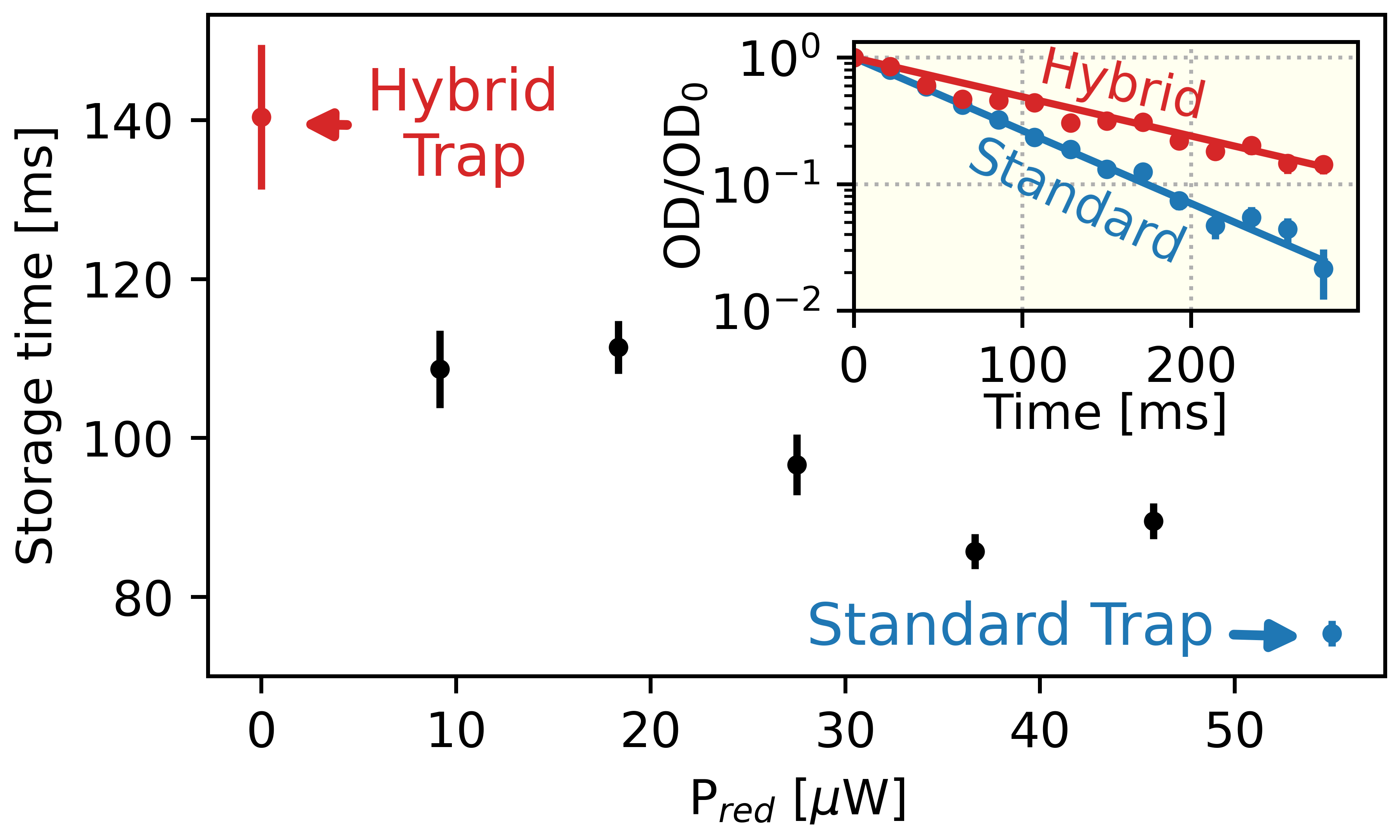}
\caption{\label{fig:FigureLifetime} Measured storage time of the trapped atoms as a function of the power of the red-detuned laser, $P_\text{red}$, during the waiting time. Our standard trap (blue dot) is obtained for $P_\text{red}$~=~55~$\mu W$, while the hybrid trap (red dot) for $P_\text{red}$ = 0 $\mu W$. Inset: normalized optical depth (OD) as a function of waiting time for the standard (blue dots and line) and hybrid (red dots and line) trap (OD$_0$ indicates the OD at the beginning of the measurement). 
}
\end{figure}

The total potential experienced by the atoms is the sum of the contributions from the trapping lasers and the surface interaction. In order to isolate and highlight the influence of the surface potential, we first load the atoms into the standard trap and then measure the OD of the ensemble after adiabatically lowering $P_{red}$ to various final values, where $P_\text{red}$ = 0 $\mu$W corresponds to the case, where the optical force is purely repulsive. The results are shown in Fig.~\ref{fig:FigurePotential}. As expected, since reducing $P_{\mathrm{red}}$ weakens the attractive trapping force, stable trapping is obtained farther away from the surface, resulting in a reduced OD. For $P_{red} > 15$~$\mu$W, the measured OD agrees well with theoretical predictions that only include the potential due to the trapping lasers (red dashed line in Fig.~\ref{fig:FigurePotential}). Accordingly, in this parameter range the surface potential plays a negligible role. For $P_{red} < 15$~$\mu$W, however, the measured OD exceeds the predicted value, suggesting that, here, the surface interaction contributes significantly to the total attractive force.

\begin{figure*}[]
\includegraphics[width=0.95\linewidth]{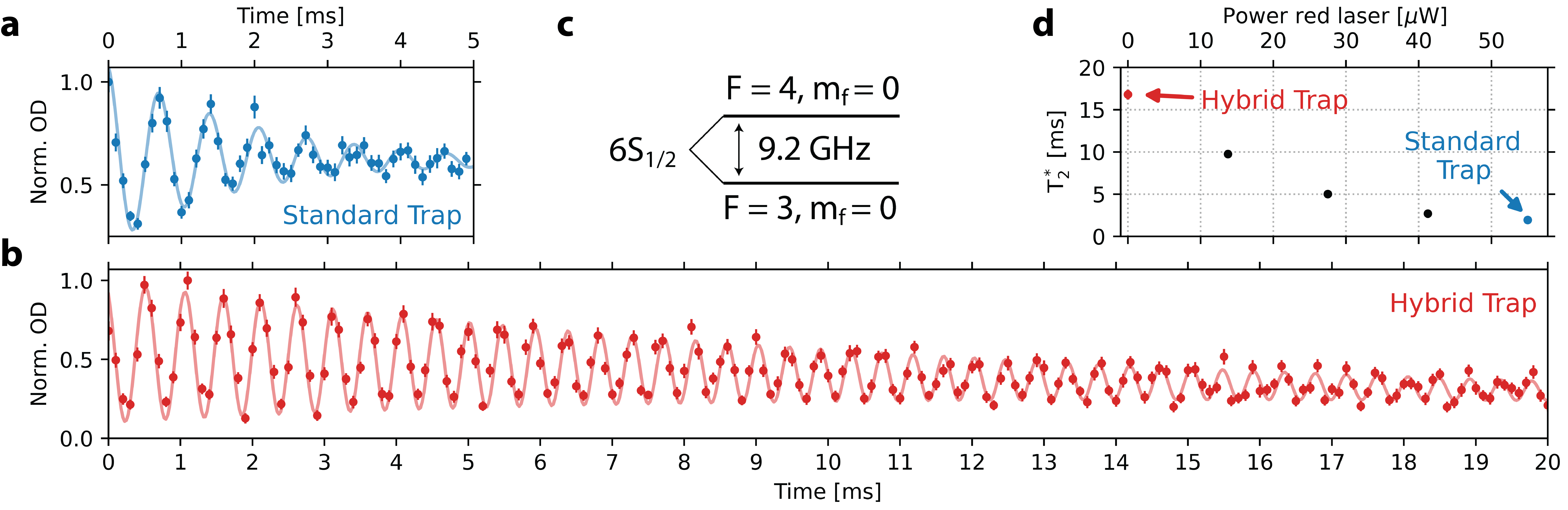}
\caption{\label{fig:FigureRamsey} Ramsey interference fringes measured on the microwave clock transition of Cs (see sketch in (c)) for (a) $P_\text{red}$=55~$\mu$W (standard trap) and (b) $P_\text{red}$=0~$\mu$W (hybrid trap). (d) Ramsey coherence time, $T_2^*$, as a function of the power of the red-detuned laser, $P_\text{red}$.
}
\end{figure*}

To interpret these findings, we develop a model which considers a surface potential that arises, to start with, solely from Casimir-Polder interactions. The Casimir-Polder potential was modeled as $U_{\mathrm{CP}}(r) = C_3/(r-r_0)^3$, with $C_3 = 1.56\,\mathrm{kHz}\,\mu\mathrm{m}^3$~\cite{Kien2007}, where, $r$ denotes the radial coordinate of the atom (with $r=0$ at the nanofiber axis) and $r_0$ the nanofiber radius. The corresponding predicted OD is shown as the dotted orange line in Fig.~\ref{fig:FigurePotential}. While this approach captures the general trend, it fails to quantitatively describe the observed data. We therefore introduced an additional attractive potential resulting from a DC Stark shift due to charges on the nanofiber surface. Assuming that these charges are uniformly distributed, this gives rise to the potential $U_{\mathrm{charges}}(r) = \frac{1}{2}\alpha_0^2\frac{\zeta^2}{4 \pi^2 \epsilon_0^2 r^2}$. Here, $\alpha_0$ is the scalar DC polarizability of Cs, $\zeta$ is the linear charge density on the nanofiber, and $\epsilon_0$ is the vacuum permittivity. The resulting prediction for the OD, shown as the solid blue line in Fig.~\ref{fig:FigurePotential}, fits our experimental data well for $\zeta = 19(3)\,\mathrm{charges}/\mu$m.
We note that, while our model captures the main features of our observations, this does not provide conclusive evidence that the surface interaction originates solely from Casimir-Polder interaction and electrostatics. However, given that our model is based on simple and reasonable assumptions, we still find it instructive to also present its predictions for further key properties of the hybrid trap, namely an estimated trap depth of $\sim$1~$\mu$K, a trap frequency of $\sim$7~kHz and a radial position of the minimum located $\sim$650~nm from the nanofiber surface.

To characterize the storage time in the hybrid trap, we first load the atoms into the conventional trap and then adiabatically lower $P_\text{red}$. After waiting for a specified duration, we transfer the remaining atoms back into the standard trap and measure the OD as a function of the waiting time. This procedure is repeated for various final values of $P_\text{red}$.
The results, shown in Fig.~\ref{fig:FigureLifetime}, reveal a counter-intuitive trend: the atom storage time increases as $P_\text{red}$ is reduced despite the trap becoming more and more shallow. In particular, in the hybrid trap, we observe a storage time of 140(9) ms, that is nearly twice as large as in the standard nanofiber trap. We note that we have repeated the measurements over several months and across a wide range of trap parameters. In all cases, the storage time in the hybrid trap consistently exceeded that of the standard trap.
We emphasize that the origin behind the excessive heating rates observed in nanophotonic atom traps compared to their free-space counterparts remains an open question \cite{Meng2018, Huemmer2019}. Given the significant differences between the hybrid and standard traps with regard to depth and the distance between the atoms and the nanofiber surface, our observation may provide critical insights for finally resolving this issue.

We now turn to the coherence properties of the atoms in the hybrid trap. Cesium atoms feature two hyperfine ground states, which can be used, e.g., to store optical information in quantum memories~\cite{Gouraud2015, Sayrin2015} or may serve as a qubit. The main figure of merit for such applications is the coherence time between these two states.
In nanophotonic systems for cold atoms, the Ramsey coherence is limited by temperature-induced inhomogeneous broadening of the clock transition~\cite{Reitz2013}. This broadening arises from motional-state-dependent differential light shifts, that in turn are proportional to the intesity of the trapping fields. As illustrated in Fig.~\ref{fig:FigureSketch}(a, b), the atoms in the hybrid trap are in a region of much lower light intensity than in the standard trap, thus, one might expect that the transfer of the atoms into the hybrid trap could result in a considerable increase of the coherence time.
To quantify the ground state coherence time in the hybrid trap, we perform Ramsey interferometry of the clock transition of Cs, i.e., $| 6 S_{1/2}, F = 3, m_f = 0 \rangle \rightarrow  | 6 S_{1/2}, F = 4, m_f = 0 \rangle$ (see Fig.~\ref{fig:FigureRamsey}), following the procedure in~\cite{Reitz2013}. In brief, we start with all atoms in the standard trap and prepare them in the state $| 6 S_{1/2}, F = 4, m_f = 0 \rangle$ by optical pumping. Then, we adiabatically lower $P_\text{red}$, launch two resonant $\pi/2$ microwave pulses separated by a certain delay time onto the atoms, transfer them back to the standard trap and then measure the resulting population in state $| 6 S_{1/2}, F = 3, m_f = 0 \rangle$. The results are shown in Fig.~\ref{fig:FigureRamsey}(a,b) for (a) $P_\text{red}$~=~55~$\mu$W (standard trap) and (b) $P_\text{red}$~=~0~$\mu$W (hybrid trap). We observe a clear increase in the Ramsey coherence time. For the standard trap, we measure $T_2^* = 1.9(1)$ ms, that is in the same order of magnitude of previous reports~\cite{Dordevic2021, Reitz2013}. The hybrid trap yields a significantly longer coherence time of $T_2^*$ = 16.8(2) ms, corresponding to an increase of about one order of magnitude. The results obtained for different final values of $P_\text{red}$ are summarized in Fig.~\ref{fig:FigureRamsey}(d) and show that the Ramsey coherence time increases as $P_\text{red}$ is reduced.
We emphasize that in these measurements we have only characterized reversible decoherence. That is, the coherence time can still be extended further using, e.g., a spin-echo sequence \cite{Hahn1950}.

In summary, we have demonstrated a new type of nanophotonic trap for laser-cooled atoms, in which surface forces are used to attract the atoms towards the waveguide, while a blue-detuned laser field provides repulsion. This strategy avoids the more involved schemes \cite{Chang2014a, Kien2022} proposed to overcome the no-go theorems associated with trapping using exclusively Casimir-Polder \cite{Rahi2010} or electrostatic forces \cite{Earnshaw1842}. Despite its shallow depth of about 1~$\mu$K, the trap can be efficiently loaded from a conventional nanophotonic trap and exhibits a remarkably long storage time. The hybrid trapping scheme opens new avenues for exploring atom–surface interactions at the nanoscale and offers distinct advantages over conventional nanophotonic traps. In particular, the ability to confine atoms in regions of low optical intensity has enabled us to achieve a new benchmark in the coherence time for quantum nanophotonic systems. This may lead to significant improvement of the performance of this platform for applications in quantum memories \cite{Simon2010, Gouraud2015, Sayrin2015, Corzo2019} and quantum information processing \cite{Chang2018, Sheremet2023, Menon2024}. In combination with the efficient loading method demonstrated here, one can envision protocols in which atoms are initially trapped close to the nanophotonic waveguide, where the high OD enables efficient storage of optical quantum information, and subsequently transferred to the hybrid trap, where enhanced coherence preserves that information. For retrieval of the stored state, the atoms could then be brought back near the waveguide surface.
Finally, our results represent an important step toward the long-sought goal of laser-free trapping of atoms around optical waveguides \cite{Chang2014a, GonzalezTudela2015a, Hung2013, Kien2022}, positioning hybrid nanophotonic traps as a powerful tool for quantum technologies.

%%%%%%%%%%%%%%%%%END OF MAIN TEXT%%%%%%%%%%%%

\subsection{Acknowledgments}

\begin{acknowledgments}
We thank Bettina Beverungen, Kurt Busch, Francesco Intravaia, and Johannes Piotrowski for helpful discussions. We acknowledge financial support by the Alexander von Humboldt Foundation in the framework of an Alexander von Humboldt Professorship endowed by the Federal Ministry of Education and Research and funding by the European Commission under the project SuperWave (ERC Grant No.101071882).
\end{acknowledgments}

\bibliography{main.bib}% Produces the bibliography via BibTeX.

\newpage
\appendix

\end{document}